\documentclass[sigconf,9pt,authorversion,nonacm]{acmart}
\makeatletter
\def\@ACM@checkaffil{%
    \if@ACM@instpresent\else
    \ClassWarningNoLine{\@classname}{No institution present for an affiliation}%
    \fi
    \if@ACM@citypresent\else
    \ClassWarningNoLine{\@classname}{No city present for an affiliation}%
    \fi
    \if@ACM@countrypresent\else
        \ClassWarningNoLine{\@classname}{No country present for an affiliation}%
    \fi
}
\makeatother
\setcitestyle{maxcitenames=1}
\usepackage{booktabs} %
\usepackage{hyperref}
\usepackage{enumitem}
\usepackage{graphicx}
\usepackage{color}
\usepackage{colortbl}
\usepackage{bm}
\usepackage{braket}
\usepackage{multirow}

\graphicspath{{figure/}{figures/}}

\acmDOI{xxx}

\acmISBN{xxx}

\begin{document}
\title{Parameter Setting Heuristics Make the Quantum Approximate Optimization Algorithm Suitable for the Early Fault-Tolerant Era}
\subtitle{(Invited Paper)}

\author{Zichang He, Ruslan Shaydulin, Dylan Herman, Changhao Li,\\ Rudy Raymond, Shree Hari Sureshbabu, Marco Pistoia}
\affiliation{%
  \institution{Global Technology Applied Research, JPMorganChase, New York, NY 10017 USA}
}

\renewcommand{\shortauthors}{Z. He et al.}

\begin{abstract}

Quantum Approximate Optimization Algorithm (QAOA) is one of the most promising quantum heuristics for combinatorial optimization.
While QAOA has been shown to perform well on small-scale instances and to provide an asymptotic speedup over state-of-the-art classical algorithms for some problems, fault-tolerance is understood to be required to realize this speedup in practice. The low resource requirements of QAOA make it particularly suitable to benchmark on early fault-tolerant quantum computing (EFTQC) hardware. However, the performance of QAOA depends crucially on the choice of the free parameters in the circuit. The task of setting these parameters is complicated in the EFTQC era by the large overheads, which preclude extensive classical optimization. In this paper, we summarize recent advances in parameter setting in QAOA and show that these advancements make EFTQC experiments with QAOA practically viable.
\end{abstract}

\ccsdesc[500]{Computer systems organization~Embedded systems}
\ccsdesc[300]{Computer systems organization~Redundancy}
\ccsdesc{Computer systems organization~Robotics}
\ccsdesc[100]{Networks~Network reliability}

\maketitle
\section{Introduction}
The Quantum Approximate Optimization Algorithm (QAOA)~\cite{farhi2014quantum, hogg2000quantum} is a quantum algorithm for solving combinatorial optimization problems. QAOA uses a series of parameterized alternating operators (``layers'') to prepare a quantum state such that its measurement outcomes correspond to high-quality solutions of the optimization problem. QAOA has attracted wide interest from the quantum computing community due to its low resource requirements and broad applicability to problems in science and industry~\cite{herman2023quantum,DAC24_review,Blekos_2024}.  
While the low resource requirements of QAOA make it possible to implement on near-term quantum devices~\cite{qaoa_np,decross2023qubit,Moses2023RaceTrack,he2023alignment,niroula2022constrained,harrigan2021quantum,Pelofske2023,Pelofske2024}, theoretical analysis suggests that achieving better-than-classical performance will likely require fault tolerance~\cite{Quek2024,DePalma2023,StilckFrana2021}.

The use of QAOA in the early fault-tolerant era is motivated by recent results providing evidence that QAOA offers an asymptotic speedup over state-of-the-art classical algorithms for certain problems~\cite{shaydulin2024evidence,boulebnane2022solving}, although with circuit sizes that requires error correction. 
As hardware progress enables increasingly sophisticated small-scale error detection and correction demonstrations~\cite{bluvstein2024logical,mayer2024benchmarking,katabarwa2024early,hangleiter2024fault}, QAOA is becoming an appealing candidate algorithm to evaluate using early fault-tolerant quantum computing (EFTQC) devices due to moderate gate count of the QAOA circuit. QAOA can be implemented for many problems with a linear or quadratic number of T gates per layer~\cite{sanders2020compilation}, making it possible to study its behavior on problems with hundreds to thousands of variables using only millions of T gates. This puts fault-tolerant QAOA within reach of early implementations of quantum error-correcting codes with modest distance, as long as QAOA depth is sufficiently small.

Though studies of QAOA in EFTQC are likely to be scientifically fruitful, we do not  expect broadly applicable speedups from such experiments in absence of significant advances in error correction. This is due to both the easiness of small-scale optimization problems and the overheads of known fault-tolerant constructions. The vast majority of industry-relevant problems that can be mapped to hundreds of qubits can be solved classically in seconds. Consequently, currently proposed quantum architectures like those based on the surface code~\cite{Babbush2021} are unlikely to be competitive in terms of time-to-solution. Nonetheless, EFTQC benchmarks are likely to provide valuable evidence on the power of QAOA to provide speedups over classical state-of-the-art, which so far have only been demonstrated on small ($\leq 40$ qubits) instances in classical simulation~\cite{shaydulin2024evidence,boulebnane2022solving}.

A central challenge in QAOA execution on hardware is the need to set the free parameters in the alternating operators. This is a challenging problem in general; in fact, it is NP-hard in the worst case~\cite{Bittel2021}. More broadly, if QAOA is viewed as a variational algorithm trained from a random initialization, it is known to suffer from the barren plateau problem in many cases~\cite{mcclean2018barren,fontana2023adjoint,Larocca2022}. 

Fortunately, a QAOA parameterized circuit is a highly structured one, enabling advanced parameter-setting techniques beyond black-box optimization techniques. Specifically, annealing-like smooth parameter schedules are both easy to optimize and offer good performance~\cite{zhou2020quantum,basso2022qaoaskmaxcut,kremenetski2021quantum,montanez2024towards}. Furthermore, QAOA parameters are known to be instance-independent if the QAOA depth (i.e. the number of alternating operators) is kept constant~\cite{brandao2018fixed,basso2022qaoaskmaxcut,2110.10685,boulebnane2022solving}. The instance-independence of QAOA parameters makes it possible to optimize the parameters for small instances from a given problem class once and use these parameters on larger instances with little or no reoptimization, enabling scaling analysis like that of Refs.~\cite{shaydulin2024evidence,boulebnane2022solving}.

The availability of good instance-independent parameters makes QAOA a viable EFTQC algorithm, in sharp contrast to variational algorithms which would require a prohibitively high number of samples to train. However, instance-specific parameter fine-tuning can still offer meaningful performance gains over directly using instance-independent parameters~\cite{Shaydulin2021}. Enabling such fine-tuning on EFTQC devices requires careful use of a limited budget of total number of quantum circuit executions or shots. While simple techniques can still perform well in this shot-limited setting under many cases~\cite{decross2023qubit,Moses2023RaceTrack,polloreno2022qaoa}, configuring the classical optimizer can lead to considerable performance improvements~\cite{hao2024end}.

In this paper, we summarize the recent progress in QAOA parameter setting, with a focus on the practical constraints associated with execution of QAOA in the early fault-tolerant era.
We show how instance-independent or fixed parameters and highly optimized parameter fine-tuning make QAOA execution possible even in the presence of EFTQC overheads. We focus in particular on Ref.~\cite{hao2024end}, which proposes an end-to-end protocol for obtaining high-quality QAOA parameters with a limited number of shots and combines many prior ideas. Our paper makes the case that QAOA is a promising heuristic for combinatorial optimization to study on EFTQC devices. We then highlight some opportunities for EFTQC experiments with QAOA, such as demonstrating its scaling advantage and benchmarking quantum hardware.

\section{Background on Quantum Approximate Optimization Algorithm}

QAOA is a quantum algorithm for solving combinatorial optimization problems. Let the $f(\bm{s})$ denote the objective function on spins $\bm{s} \in {\{-1,1\}}^N$ to be optimized. %
A commonly studied class of objective functions can be written as a polynomial on spins 
as follows,
\begin{equation}\label{eq:classical_cost}
 f(\bm{s}) = \sum_{\{u_1,\ldots, u_k\}} w^{(k)}_{u_1\ldots u_k}s_{u_1}\ldots s_{u_k} + \ldots + \sum_{u} w^{(1)}_{u} s_u.
\end{equation}
QAOA solves optimization problems by applying in alternation a problem-specific Hamiltonian, which encodes the optimization problem, and a mixing Hamiltonian, which mixes together probability amplitudes and promotes exploration of the solution space. The circuit is defined by two operators, problem Hamiltonian $\bm{H}_P$ which is mapped from the classical cost function $f(\bm{s})$ and mixer Hamiltonian $\bm{H}_M$, a hyperparameter $p$, and an initial state $\ket{\bm{\psi}_0}$:
\begin{align}
\label{eqn:qaoa_state}
    \left\vert {{{\bm{\psi }}}}({{{\bm{\gamma }}}},{{{\bm{\beta }}}})\right\rangle ={e}^{-i{\beta }_{p}{{{{\bm{H}}}}}_{M}}{e}^{-i{\gamma }_{p}{{{{\bm{H}}}}}_{P}}\ldots {e}^{-i{\beta }_{1}{{{{\bm{H}}}}}_{M}}{e}^{-i{\gamma }_{1}{{{{\bm{H}}}}}_{P}}\left\vert {{{{\bm{\psi }}}}}_{0}\right\rangle ,
\end{align}
where $\bm{\gamma}$ and $\bm{\beta}$ are parameters associated with these $2p$ number of operators. The algorithm's performance is controlled by the quality of these parameters, which can be set analytically or optimized by classical techniques. A typical problem formulation for optimizing these parameters is to minimize the energy
\begin{equation}
    \min_{\bm{\gamma}, \bm{\beta}} \braket{\bm{\psi}(\bm{\gamma}, \bm{\beta})|\bm{H}_P|\bm{\psi}(\bm{\gamma}, \bm{\beta})}.
\end{equation}
While there are many variants of the QAOA~\cite{Hadfield_2019}, unless specified otherwise, we take the original form~\cite{hogg2000quantum} 
whose initial state is a $\ket{\bm{+}}$ state and mixer ${\bm{H}}_{M} = \sum_i \bm{X}_i$ throughout this paper. Let $f_{\rm{min}}$ and $f_{\rm{max}}$ be the minimum and maximum value of the classical cost function. The solution quality of a QAOA state is quantified by the approximation ratio (AR): 
\begin{equation}
    AR = \frac{\braket{\bm{\psi}(\bm{\gamma}, \bm{\beta})|\bm{H}_P|\bm{\psi}(\bm{\gamma}, \bm{\beta})} - f_{\rm{max}}}{f_{\rm{min}} - f_{\rm{max}}}.
\end{equation}

\subsection{Properties of QAOA parameterized quantum circuit and connections to other algorithms}

One can view QAOA as a variational quantum eigensolver (VQE) \cite{peruzzo2014variational} with the specific parameterized quantum circuit or ``ansatz'' given by Eq.~\eqref{eqn:qaoa_state}, leading to a structured cost landscape~\cite{hao2024variational,stkechly2023connecting}. 
In this framework, the alternating operator ansatz is a problem-specific ansatz that possesses a few intuitive interpretations as well as desirable properties, some theoretical and some empirically observed.

One interpretation of the alternating operator ansatz is that the mixer performs a continuous-time quantum walk on the feasible subspace and the cost operator applies distinguishing phases according to the bistring's cost, analogous to the marking operation in Grover's algorithm for search problems. Specifically, quantum unstructured search can viewed in the singular value transformation framework \cite{gilyen2019quantum} as a sequence of similar unitaries as in the QAOA, where the mixer generates a walk over the complete graph, and the angles are such that the initial state is driven towards the target state.

Another interpretation of QAOA that is closer in spirit to the task at hand, optimization, is the connection to the adiabatic quantum optimization (AQO) algorithm \cite{farhi2000quantum}. If the unitaries at each layer of QAOA are changing sufficiently slowly, the discrete-time adiabatic theorem \cite{dranov1998discrete, costa2022optimal} states that if the initial state is an eigenstate of $H_{M}$, then QAOA will evolve the state along a path connecting corresponding eigenstates of the QAOA unitary at each layer \cite{kremenetski2023slowchanginuitaries}. Unfortunately, in the slow-changing unitaries regime, the final state may not be the ground state of $H_c$ due to eigenphase wrap around. Alternatively, if the angles are also vanishingly small, removing Trotter error, then we approach the continuous-time adiabatic regime, which will produce the desired ground state if the evolution is sufficiently slow \cite{farhi2014quantum}. Additionally, the Trotter error in QAOA can also be viewed as an approximate counter-diabatic (CD) term \cite{Kolodrubetz2017} potentially enabling accelerated evolution over AQO \cite{Wurtz_2022}. Such CD phenomena is important as AQO, due the minimum spectral gap, can have a runtime  that is super exponential \cite{altshuler2010anderson}.

\section{Overview of QAOA parameter setting methods}\label{sec:para_setting}
Optimizing the parameters $\bm{\gamma}, \bm{\beta}$ is a crucial step in QAOA execution and is a prerequisite for achieving any potential quantum speedup. While optimal QAOA parameters can be derived analytically for some problems in small constant depth ($p\leq 2$) by analyzing the trigonometric formula for the objective function~\cite{hadfield2018quantumalgorithmsscientificcomputing, Wang_2018,1912.12277,sureshbabu2024parameter}, theoretical analysis is challenging for larger $p$. This is an important challenge since large $p$ is likely required for quantum advantage~\cite{farhi2020quantumapproximateoptimizationalgorithm, PhysRevLett.125.260505}. 

To overcome this challenge, many techniques have been developed in the recent years, which make it possible to execute QAOA without expensive parameter optimization. These techniques make QAOA an experimentally viable EFTQC algorithm. We now overview some of these techniques. We discuss an end-to-end pipeline combining some of these in Sec.~\ref{sec:end-to-end} below.

\subsection{Parameter concentration and transferability}~\label{sec:para_concentration}

Parameter concentration refers to the optimal parameters being concentrated over random instances from a given problem class. For example, if we consider solving MaxCut on two random $d$-regular graphs, with high probability, the optimal QAOA parameters for both of them will be close. This enables parameter transferability: if optimal parameters for instances of a given problem class are all close to each other, optimized parameters for one instance can be transferred to another.

The fundamental fact underlying this phenomenon is the concentration of the QAOA energy if QAOA depth is constant. For various problems, the QAOA state has been shown to have a mean energy that concentrates over random problem instances \cite{Farhi_2022SK, brandao2018fixed, basso2022qaoaskmaxcut,sureshbabu2024parameter} and over measurements \cite{farhi2014quantum, Farhi_2022SK}. The former implies performance of the algorithm achieves  a ``typical'' value with high probability and the later implies that energy of the QAOA state with respect the cost function is a good indicator of actual performance. Interestingly, the concentration phenomenon of constant-depth QAOA over MaxCut instances on high-girth, $d$-regular graphs \cite{basso2022qaoaskmaxcut, sureshbabu2024parameter} presents a setting where the cost function attains a typical value over problems with different numbers of variables. This hints that optimized parameters for small instances could transfer and perform similarly on larger instances. Unfortunately, in the $p=o(\log(n))$ regime, the smoothness of QAOA over random instances also leads it to suffer from topological barriers that plague classical local approximation algorithms applied to hard combinatorial problems \cite{farhi2020quantumogp}. 

The previously mentioned results on QAOA parameter transferability were focused on average case performance, where the goal is to use QAOA as an approximation algorithm, running in polynomial time. It is also possible to utilize QAOA  as an exact solver, i.e. the probability of measuring the optimal solution is the target metric \cite{boulebnane2022solving, shaydulin2024evidence}. In this setting, we are comparing to algorithms that run in exponential time, and thus the previously mentioned topological barriers are no longer relevant. For example, one work showed that, for the Low Autocorrelation Binary Sequences (LABS) problem, constant-depth QAOA with a fixed set of parameters, obtained at small problems sizes, produces a decay in overlap that, when combined with quantum minimum finding \cite{shaydulin2024evidence}, outperformed the best classical exact solutions, i.e. a scaling advantage. This is especially interesting given that LABS has only one problem instance for a given number of variables. A similar analysis was also done for random $k$-SAT \cite{boulebnane2022solving}. The transferability of QAOA parameters optimized with respect to ground state fidelity still lacks a theoretical explanation as the standard concentration arguments of e.g. \cite{Farhi_2022SK, basso2022qaoaskmaxcut} do not apply.

\subsection{Schedule reparameterization}\label{sec:para_schedule}
When QAOA depth $p$ is large, directly optimizing the $2p$ parameters $\bm{\gamma}, \bm{\beta}$ often becomes impractical. To address this challenge, multiple reparameterizations of QAOA parameters have been introduced. Specifically, in lieu of optimizing $\bm{\gamma}, \bm{\beta}$ directly, one can optimize a different (potentially, smaller or better-behaved) set of parameters that are connected to $\bm{\gamma}, \bm{\beta}$ by some fixed transformation. We now discuss a few such parameterizations.

\begin{enumerate}[leftmargin=*]

    \item \textbf{Linear Ramp}: The simplest parameterization is inspired by quantum annealing and corresponds to linearly ``ramping up'' the cost Hamiltonian and ``ramping down'' the mixer~\cite{Shaydulin2021Classical,kremenetski2021quantum,Sack2021}: 
    \begin{align}
        \gamma(f_j) &= \Delta f_j \label{lin_gamma}
        \\
        \beta(f_j) &= \Delta (1-f_j) \label{lin_beta}
        \\
        f_j &= \frac{j}{p+1},
    \end{align}
    where $f \in [0,1]$ with $j=1,2,...,p$ and $\Delta \in \mathbb{R}$ is a parameter to tune.
    This schedule can be extended to include different constants for $\bm{\gamma}$ and $\bm{\beta}$ \cite{Shaydulin2021Classical,montanez2024towards}:
    \begin{align}
        \gamma(f_j) &= \Delta^{(1)}_{\gamma} f_j + \Delta^{(2)}_{\gamma}
        \\
        \beta(f_j) &= \Delta^{(1)}_{\beta} (1-f_j) + \Delta^{(2)}_{\beta},
    \end{align}
    where $\Delta^{(1)}_{\gamma}, \Delta^{(2)}_{\gamma}$ and $\Delta^{(1)}_{\beta}, \Delta^{(2)}_{\beta}$ can be tuned individually to obtain the best performance.
    \\
    A linear interpolation approach, termed INTERP can be used to extend the optimized $p-1$ layer parameters to initialize the $p$ layer parameters and an optimization is performed at the $p$ layer~\cite{zhou2020quantum}:
    \begin{align}
        {\gamma_{\rm{init}}}^{(p+1)}_{j} &= \frac{j-1}{p} \gamma^{(p)}_{j-1} + \frac{p-j+1}{p} \gamma^{(p)}_{j}\\
        {\beta_{\rm{init}}}^{(p+1)}_{j} &= \frac{j-1}{p} \beta^{(p)}_{j-1} + \frac{p-j+1}{p} \beta^{(p)}_{j},
    \end{align}
    where $\gamma^{(p)}_{j}$ and $\beta^{(p)}_{j}$ are the $j$-th parameters optimized at $p$ layer. 
    
    \item \textbf{Non-linear}:
    \begin{itemize}
        \item \textbf{Root}: Having noticed that the smaller gaps occur toward the end of the evolution in certain chemistry problems, a \textbf{root} schedule was introduced such that smaller increments can happen at the end \cite{kremenetski2021quantum}. This schedule is obtained by replacing $f$ in Eq. \eqref{lin_gamma} and Eq. \eqref{lin_beta} by $\sqrt{f}$.

        \item \textbf{Tangent}: A tangent schedule uses smaller steps in the middle and larger increments at the beginning and toward the end of the evolution \cite{kremenetski2021quantum}. This schedule is obtained by replacing $f$ in Eq. \eqref{lin_gamma} and Eq. \eqref{lin_beta} by $\frac{1}{2\tan\left(\frac{1}{2c}\right)}\left( \tan\left(\frac{f-0.5}{c}\right) + \tan\left(\frac{1}{2c}\right) \right)$, with $c$ being a constant. Ref.~\cite{kremenetski2021quantum} notes that \textbf{root} and \textbf{tangent} schedules perform similarly to a linear ramp.

        \item \textbf{Fourier}: The discrete $\sin$ and $\cos$ transforms can be utilized to determine the parameters at the $p$-th layer from the optimized \textbf{fourier} amplitudes from the $p-1$ layer \cite{zhou2020quantum}, by accounting for $q$ frequency amplitudes. This can be written in the form:
        \begin{align}
            \gamma_j &= \sum_{k=1}^{q} u_k sin\left[\left(k-\frac{1}{2}\right)\left(j-\frac{1}{2}\right)\frac{\pi}{p}\right]
            \\
            \beta_j &= \sum_{k=1}^{q} v_k cos\left[\left(k-\frac{1}{2}\right)\left(j-\frac{1}{2}\right)\frac{\pi}{p}\right],
        \end{align}
        where $u_k$ and $v_k$ are coefficients to optimize.
    \end{itemize}

\end{enumerate}

\subsection{Parameter optimization}
While parameter schedule, concentration, and transferability help obtain a good initial parameter set, it is often the case that the parameters must be fine-tuned to achieve better accuracy. This requires the use of performant local optimizers. While gradient-based optimizers are popular in classical machine learning due to the low cost of gradient evaluation enabled by backpropagation, the best known techniques for obtaining gradients of parameterized quantum circuits have cost that grows with the number of parameters~\cite{10.5555/3666122.3668062}. The resulting high sample cost of evaluating gradients makes gradient-based optimization methods largely uncompetitive. Preference is therefore given to gradient-free optimizers, typically but not exclusively trust-region ones. 
Examples of optimizer benchmarks include Refs.~\cite{Shaydulin2019,Nannicini2019,bonet2023performance,1701.01450,hao2024end}.

Parameter or QAOA landscape concentration can be exploited by learning-based optimizers, which can ``learn'' the instance-independent landscape and optimize over it~\cite{khairy2020learning, pmlr-v107-yao20a, Wauters_2020, cheng2024quantum, he2024_dro}. While such techniques can achieve good performance, it is unclear if they can outperform simpler parameter transfer.

\section{End-to-End Parameter Setting Protocol}\label{sec:end-to-end}
\begin{figure}[ht!]
  \includegraphics[width = \linewidth]{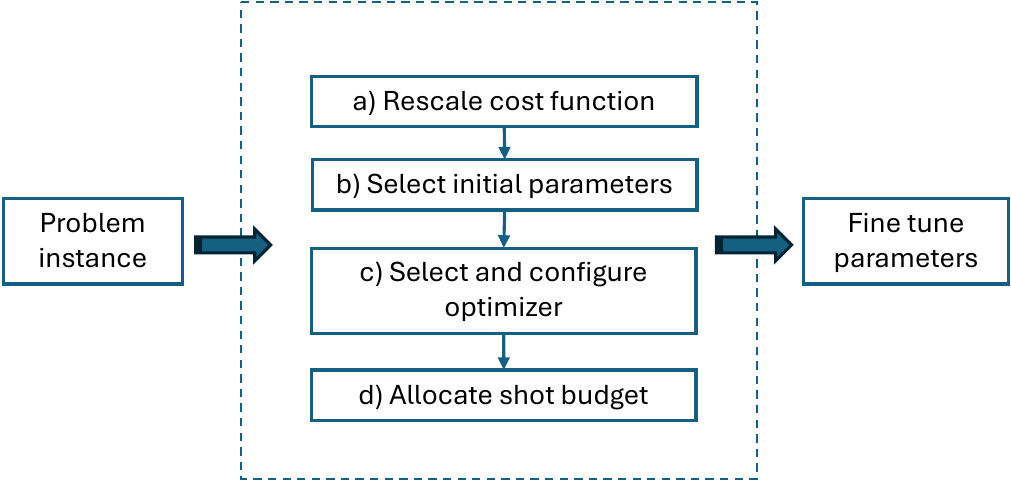}
  \caption{Overview of the end-to-end parameter setting protocol. 
  } 
  \label{fig:protocol}
\end{figure}
Unifying the discussed parameter setting approaches above, Ref.~\cite{hao2024end}, proposes an end-to-end protocol for setting high-quality QAOA parameter for the execution on a quantum processor, as summarized in Fig.~\ref{fig:protocol}. The key idea is to start with good instance-independent initial parameters and fine tune the parameters per instance. Since fine-tuning can be accomplished with a small ($\leq 10$k) shot budget, this end-to-end protocol is directly applicable to the shot-constrained EFTQC setting. Ref.~\cite{hao2024end} shows that the protocol can tolerate realistic hardware noise by performing executions of QAOA with $N=32$ and $p=5$, using up to 240 two-qubit gates.

We now describe protocol step-by-step.

\textbf{a) Rescaling the problem cost funciton.} 
This is a pre-processing step that is extremely simple but useful for weighed problems. Given a problem defined as Eq.~\eqref{eq:classical_cost}, divide the original cost function by the following scaling factor:
\begin{equation}
    \sqrt{\frac{1}{\lvert E_{k}\rvert}\sum_{\{u_{1},\dots, u_{k}\}}(w^{(k)}_{u_1,\dots, u_k})^2 + \ldots + \frac{1}{\lvert E_1\rvert}\sum_{u}(w^{(1)}_{u})^2}
\end{equation}
where each $|E_j|$ is the number of terms of order $j$.
Applying such a constant scalar does not change the optimium solution of the original problem, but it will make the parameter optimization much easier~\cite{sureshbabu2024parameter}. 
This scaling rule has been shown theoretically to be optimal for weighted MaxCut on large-girth regular graphs with i.i.d. weights~\cite{sureshbabu2024parameter}, and performs well empirically for a broad class of problems.

\textbf{b) Selecting initial parameter.} The initial parameters significantly impact the quality of optimized parameters. It has been shown for many problem settings that the optimal parameters of different problem instances are very similar~\cite{basso2022qaoaskmaxcut,shaydulinParameterTransferQuantum2023,sureshbabu2024parameter,shaydulin2024evidence,boulebnane2022solving,2110.10685}. Consequently, the averaged optimized parameters from several problem instances serve as a high-quality initial point. Beyond the concentrated parameters, other initial parameter setting strategy discussed in Sec.~\ref{sec:para_setting} could be adopted as well. 

\textbf{c) Selecting and configuring a classical optimizer.}
Although for most problems there exists an initial parameter set that works well for many instances, the instance-level QAOA performance can be further improved by fine tuning leveraging a classical optimizer. Ref.~\cite{hao2024end} performs an extensive benchmark of optimizers in the EFTQC-relevant shot-frugal setting (total budget of $10,000$ shots to optimize the parameters). 
The benchmark shows that trust-region optimizers with a simple internal model, namely COBYLA~\cite{COBYLA} (linear model) and BOBYQA~\cite{BOBYQA} (quadratic model) perform best in this setting. This is due to simple models requiring only a small initial stencil to calibrate an initial model at the start of optimization.

For such techniques, one of the most important hyperparameters is the initial step size to build and update the model, called ``rhobeg'' in COBYLA's scipy implementation. Ref.~\cite{hao2024end} shows that COBYLA's performance is not very sensitive to initial step size, though smaller values are preferable if the quality of initial parameter guess is higher.

\textbf{d) Allocating shot budget.}
Given a limited shot budget, an important question is how to distribute it over many iterations. For example, one option is to run the optimizer for many steps and each step allocate only a small shot budget; alternatively, optimizer can be run for only a few steps with each step allotted a bigger budget. Ref.~\cite{hao2024end} finds that, 
given a total budget of $10,000$ shots and availability of a good initialization, the most performant strategy is to take only two additional steps after the initial stencil.

\section{Opportunities for EFTQC experiments with QAOA}\label{sec:applications}
The relative ease of compilation~\cite{sanders2020compilation,10.5555/3179330.3179331} and the availability of good parameter-setting procedures make QAOA a promising candidate algorithm to evaluate on EFTQC devices. We now briefly discuss promising opportunities for such experiments.

\subsection{Evaluating the evidence of QAOA speedup}
\begin{table}[t]
\begin{tabular}{cc|cc|c}
\hline
\multicolumn{2}{c|}{Problem}   & QAOA & QAOA$+$QMF                                              & Best classical        \\ \hline
\multirow{2}{*}{LABS}  & Fit   & 0.546 (p=12)        & \textbf{0.275} (p=12)          & 0.432                          \\
                       & CI    & [0.506,0.585]     & [0.253,0.293]                 & [0.411,0.465]  \\ \hline
\multirow{2}{*}{8-SAT} & Fit   & 0.302 (p=60) & \textbf{0.151} (p=60) & 0.325                   \\
                       & CI & [0.295, 0.309]   & [0.130,0.155]                             & [0.317, 0.333]                         \\ \hline
\end{tabular}
\caption{Summary of evidence of scaling advantage of QAOA. The fitting of time-to-solution over problem size $N$ follows the formula $2^{\alpha N} + c$. The table reports the fitted scaling factor $\alpha$ and its $95\%$ confidence intervals (CI). The data for LABS problem is from Ref.~\cite{shaydulin2024evidence}, and the data for 8-SAT problem is from Ref.~\cite{boulebnane2022solving}. ``QAOA$+$QMF'' refers to QAOA combined with quantum minimum finding (QMF), which offers a quadratically better performance than QAOA alone. ``Best classical'' includes both exact solvers and empirically evaluated heuristics.
}\label{tab:summary}
\end{table}

The first and most clear opportunity is validating and extending the recent results observing a quantum speedup with QAOA. Numerical evidence of a quantum speedup using QAOA has been demonstrated over state-of-the-art classical solvers for the Low Autocorrelation Binary Sequences (LABS) problem~\cite{shaydulin2024evidence} and random $k$-SAT~\cite{boulebnane2022solving}, with further numerical evidence available for good scaling on multiple combinatorial optimization problems~\cite{montanez2024towards}.

These results are obtained using QAOA with fixed parameters. Specifically, Refs.~\cite{shaydulin2024evidence,boulebnane2022solving} set the fixed parameters to be mean QAOA parameters over appropriate rescaled parameters optimized for small problem instances. Rescaled fixed parameters are then used for larger problem instances. Ref.~\cite{montanez2024towards} uses fixed linear ramps, with ramp parameters chosen heuristically.

The results of Refs.~\cite{shaydulin2024evidence,boulebnane2022solving,montanez2024towards} have two important limitations which can be overcome by EFTQC experimentation. First, they focus on small problems. This is due to the results obtained primarily by classical simulation of deep circuits with exponential precision required to accurately characterize the scaling. In this regime, the most performant simulation technique is full statevector propagation, which can only be achieved for a modest number of qubits ($\leq 40$ in \cite{shaydulin2024evidence} and $\leq 42$ in \cite{montanez2024towards}). Second, as a consequence of the small qubit count, the scaling can only be determined for small $p$. This is because for high $p$ but low-qubit count, the overlap gets too large, making the scaling estimated from small $p$ unrepresentative. Experiments on EFTQC devices using hundreds of qubits would allow exploration of the higher-depth regime, which has the potential to enable better scaling~\cite{boulebnane2022solving}.

\subsection{Benchmarking quantum hardware}

The quality of QAOA execution on an imperfect quantum device can be esaily measured by QAOA energy, making it a convenient benchmark to evaluate and compare quantum devices. In fact, execution of QAOA with a large product of qubit count and depth was one of the stated goals of DARPA Optimization with Noisy Intermediate-Scale Quantum program~\cite{DARPA_ONISQ_BAA}. As QAOA can be executed with arbitrary depth on hardware, a meaningful benchmark is the product between the qubit count $N$ and the largest depth $p$ such that QAOA performance with depth $p$ is better than with depth $p-1$ when executed on hardware using $N$ qubits.

Ref.~\cite{qaoa_np} executes this benchmark on trapped-ion quantum processors, achieving $N\cdot p$ of up to 320 using $N=32$ and $p\leq 10$. Ref.~\cite{Pelofske2024} shows $N \cdot p$ of up to $127\cdot 2 = 254$ on superconducting quantum processors. To the best of our knowledge, these results remain the state-of-the-art, though they are likely to be improved on the next-generation quantum hardware currently available on the cloud.

As hardware continues to progress, moving to larger $N\cdot p$ would require error correction. QAOA offers a holistic benchmark that enabled comparing different codes, fault-tolerant architectures and compilation strategies.

\section{Discussion}
As this paper shows, recent advances largely solve the problem of parameter setting in QAOA with constant depth. Combination of good initial instance-independent parameters and shot-frugal fine-tuning achieves good performance, primarily due to parameter concentration enabling very high-quality initial guess. 
While schedule reparameterization techniques (e.g., linear ramp) take a step towards addressing the challenge of running QAOA in high depth, their performance is sometimes not as expected and could be drastically worse than direct QAOA parameter optimization. 
This is in sharp contrast to the small constant depth regime, where instance-independent parameters are typically nearly optimal. The development of parameter-setting techniques that work beyond small or constant depth is therefore an important challenge for the community.

Even in absence of such advances, current techniques enable EFTQC experimentation with small (e.g.,, $p < 50$) constant depth QAOA. As we outline in Sec.~\ref{sec:applications}, these are likely to yield valuable insights both into the potential of QAOA to provide practically-relevant speedups and into the relative strengths of quantum hardware. This also paves the way for observing interesting insights into the algortihmic aspects of QAOA.

\bibliographystyle{abbrv}
\bibliography{Bib/reference.bib} 

\section*{Disclaimer}
This paper was prepared for informational purposes by the Global Technology Applied Research center of JPMorgan Chase \& Co. This paper is not a product of the Research Department of JPMorgan Chase \& Co. or its affiliates. Neither JPMorgan Chase \& Co. nor any of its affiliates makes any explicit or implied representation or warranty and none of them accept any liability in connection with this paper, including, without limitation, with respect to the completeness, accuracy, or reliability of the information contained herein and the potential legal, compliance, tax, or accounting effects thereof. This document is not intended as investment research or investment advice, or as a recommendation, offer, or solicitation for the purchase or sale of any security, financial instrument, financial product or service, or to be used in any way for evaluating the merits of participating in any transaction.

\end{document}